\documentclass[aip,apl,reprint]{revtex4-1}

\usepackage{graphicx}% Include figure files
\usepackage{amsmath}

\begin{document}

%\preprint{}

\title{Spin-wave propagation and transformation in a thermal gradient} %Title of paper



\author{Bj{\"o}rn~Obry, Vitaliy~I.~Vasyuchka, Andrii~V.~Chumak, Alexander~A.~Serga, and Burkard~Hillebrands}
\noaffiliation
%\author{Vitaliy~I.~Vasyuchka}
%\author{Andrii~V.~Chumak}
%\author{Alexander~A.~Serga}
%\author{Burkard~Hillebrands}
%\thanks{}
\affiliation{Fachbereich Physik and Forschungszentrum OPTIMAS, Technische Universit\"at Kaiserslautern, D-67663 Kaiserslautern, Germany}


\date{\today}

\begin{abstract}
The influence of a thermal gradient on the propagation properties of externally excited dipolar spin waves in a magnetic insulator waveguide is investigated. It is shown that spin waves propagating towards a colder region along the magnetization direction continuously reduce their wavelength. The wavelength increase of a wave propagating into a hotter region was utilized to realize its decomposition in the partial waveguide modes which are reflected at different locations. This influence of temperature on spin-wave properties is mainly caused by a change in the saturation magnetization and yields promising opportunities for the manipulation of spin waves in spin-caloritronic applications.
\end{abstract}


\maketitle %\maketitle must follow title, authors, abstract and \pacs

The rapidly evolving field of spin caloritronics\cite{Bauer2012} has benefited from the discovery of the spin Seebeck effect in magnetic insulators\cite{Uchida2010}. Due to the spin Seebeck effect a heat flow gives rise to a spin current which can be converted into a charge current utilizing the inverse spin Hall effect (ISHE)\cite{Hirsch1999}. Since the flow of charges is prohibited in a magnetic insulator, the thermally induced generation of an ISHE voltage can be attributed to spin waves\cite{Uchida2010,Uchida2010a,Xiao2010}.

This opens a set of fundamental questions on the interplay between temperature and spin waves, since temperature is strongly coupled to magnetic properties like saturation magnetization, anisotropy fields and exchange constant. In particular, the behavior of an externally excited, coherent spin wave in a thermal gradient, i.e. in a region of continuously varying temperature, has not yet been studied.

In this letter we present investigations on externally excited spin waves which propagate along a magnetic insulator spin-wave waveguide with varying temperature. It is shown that a thermal gradient along the direction of propagation yields a continuous change in the spin-wave wavelength with the local waveguide temperature. In the case of a propagation towards a hotter region a reflection of spin waves is observed, since their existence in a region above a critical temperature is prohibited. The results show that a controlled manipulation of spin-wave propagation by means of temperature is possible.

\begin{figure}
	\includegraphics[viewport = 84 442 445 587, clip, scale=1.0, width=1.0\columnwidth]{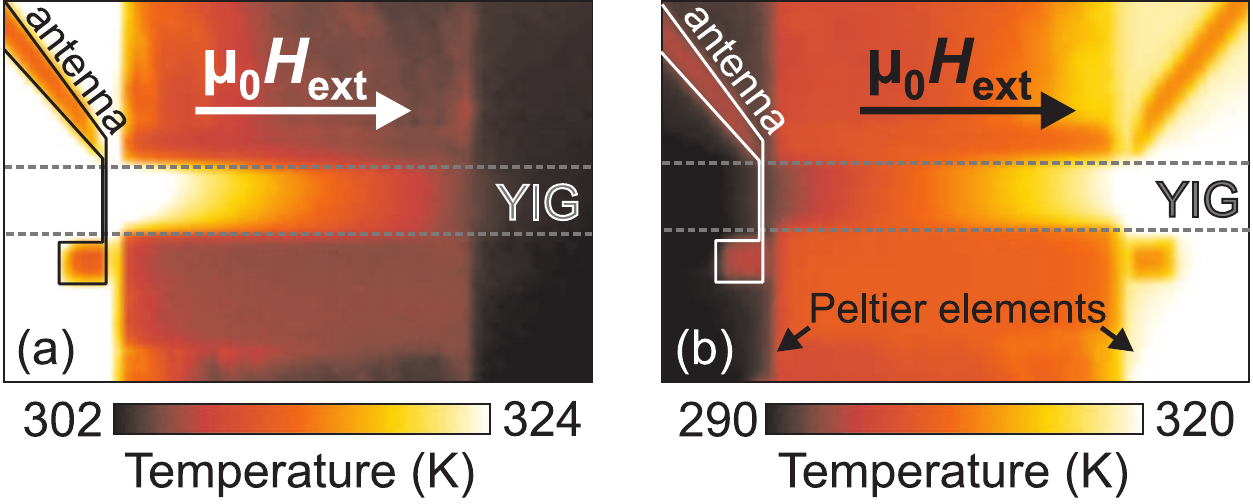}
	\caption{\label{Fig1} (Color online) Thermal images of the sample with overlaid schematic setup. A microwave antenna excites spin waves in the YIG waveguide, which then propagate along the thermal gradient towards a colder (a) and a hotter (b) region and parallel to a static external field $\mu_\text{0}H_{\text{ext}}$. The temperature difference between the left and right side of the waveguide is created by two Peltier elements.}
\end{figure}

In the experiment spin waves have been excited in an yttrium iron garnet (YIG) waveguide with a width of $w = 1.5$\,mm and a thickness of $t = 6.7\,\mu$m (indicated by the dashed lines in Fig.\,\ref{Fig1}). In order to avoid reflections of spin waves from the ends of the waveguide, a $28$\,mm long stripe with beveled edges has been used. Each end of the stripe is mounted on a Peltier element allowing for the application of a temperature gradient along the long axis of the stripe. Figure\,\ref{Fig1} shows thermal images of the experimental setup for decreasing (Fig.\,\ref{Fig1}(a)) and increasing (Fig.\,\ref{Fig1}(b)) temperature along the waveguide, respectively. Spin waves are excited by applying a microwave current to a lithographically produced microstrip line with a width of $50\,\mu$m serving as the antenna and allowing for the excitation of dipolar dominated spin waves (accessible wave vector $k\leq 60$\,rad/mm). Due to a static external magnetic field $\mu_0 H_{\text{ext}}$ applied parallel to the stripe's long axis backward volume magnetostatic spin waves (BVMSWs)\cite{Damon1961} are excited. The spin waves are investigated using Brillouin light scattering spectroscopy (BLS), which is based on the inelastic scattering of light from spin waves\cite{Hillebrands2000} providing a high spatial resolution of $50\,\mu$m and a spectral resolution of $250$\,MHz.

\begin{figure}
	\includegraphics[viewport = 205 119 556 452, clip, scale=1.0, width=0.9\columnwidth]{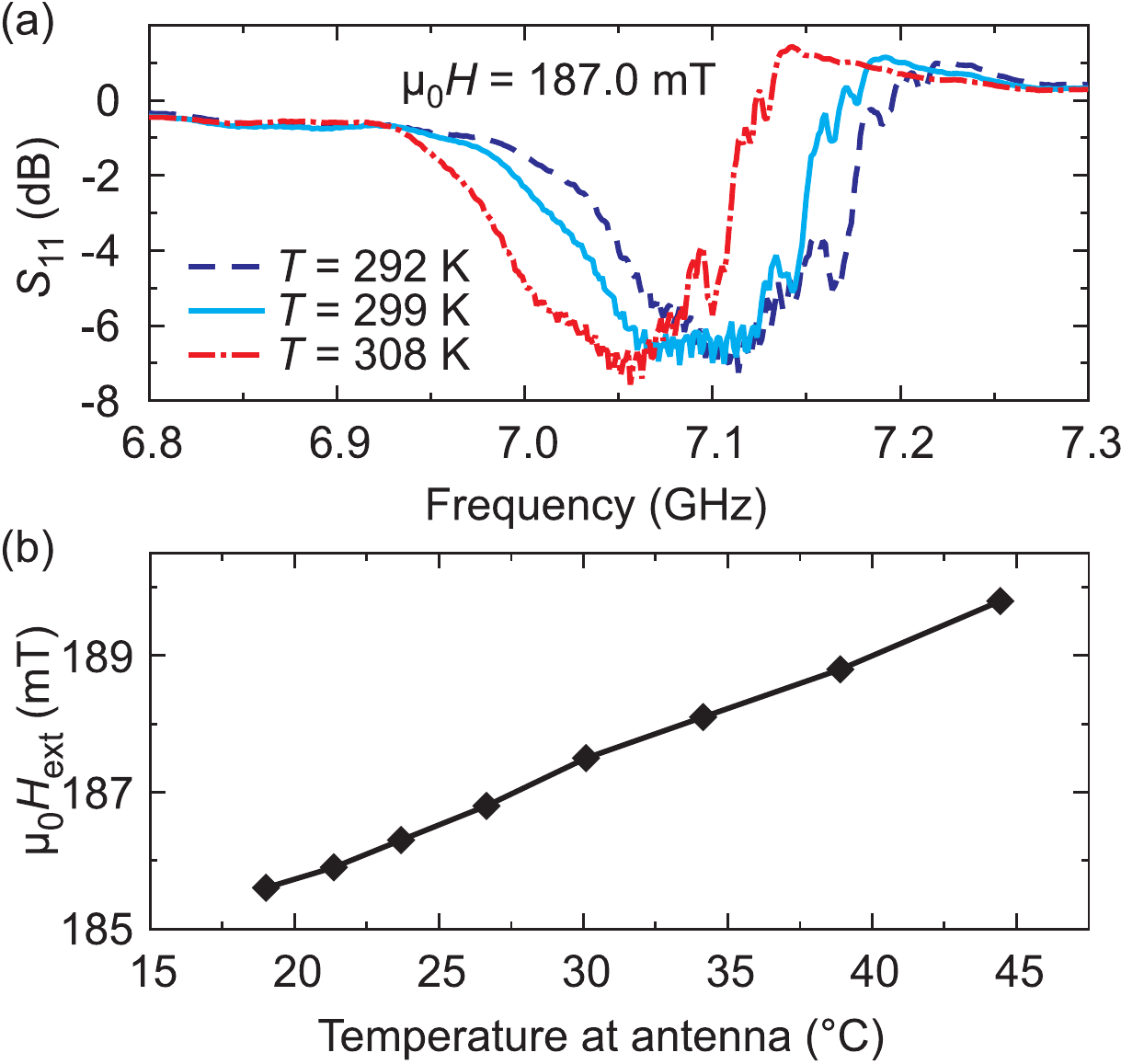}
	\caption{\label{Fig2} (Color online) Temperature dependence of the spin-wave spectrum. (a) Scalar network analyzer measurements of the microwave signal being reflected at the excitation antenna for an external field of $\mu_\text{0}H_{\text{ext}} = 187.0$\,mT and different temperatures. A minimum in the $S_\text{11}$ parameter indicates maximum absorption by the YIG stripe. (b) Dependence of the external magnetic field which is needed to compensate the dispersion shift in (a) on the temperature measured at the position of the antenna.}
\end{figure}

As a first step the dependence of the spin-wave excitation spectrum on the local temperature at the antenna is determined for an external field of $\mu_\text{0}H_{\text{ext}} = 187.0$\,mT. For this purpose, a microwave current is applied to the excitation antenna and the reflected signal ($S_\text{11}$ parameter) is detected by a network analyzer (Wiltron 54161A). The respective temperature of the YIG stripe near the antenna has been measured with an infrared camera (FLIR SC655) with a thermal sensitivity of $50$\,mK. The results are shown in Fig.\,\ref{Fig2}(a). A minimum in the $S_\text{11}$ parameter indicates maximum microwave energy absorption by the YIG stripe. With increasing temperature a shift of the absorption curve to lower frequencies is observed. This is attributed to a heat induced change in the magnetic properties of the YIG, which affects the spin-wave dispersion relation. In the simplest case of ferromagnetic resonance (FMR) excitation, i.e. all spins precess in phase, the dispersion is given by Kittel's formula\cite{Kittel1948}

\begin{equation}
  \label{Eq1}
	\nu_\text{FMR}(T) = \mu_0\gamma\sqrt{H_\text{ext}\left[H_\text{ext}+M_\text{S}(T)+H_\text{a}(T)\right]},
\end{equation}
with the gyromagnetic ratio $\gamma = 28$\,GHz/T, the external magnetic field $H_\text{ext}$, the saturation magnetization $M_\text{S}$ and the anisotropy field $H_\text{a}$\cite{Gieniusz1987,Kaack1999}. Here, $M_\text{S}$ and $H_\text{a}$ depend on the temperature and cause the frequency shift in the resonance spectrum. For a homogeneous magnet below the Curie temperature $T_\text{C}$, the saturation magnetization can be obtained by averaging the total magnetic moment over the magnetic volume. $M_\text{S}(T)$ can be calculated from the molecular field theory\cite{Hansen1974,Roeschmann1981}. For $T \lesssim 2/3\cdot T_\text{C}$, however, the temperature dependence is better described by the contribution of thermally activated spin waves\cite{Gurevich1996,Lvov1993}. As a consequence, heating the magnet reduces its saturation magnetization. According to Eq.\,\ref{Eq1} this results in a reduction of the spin-wave frequency. An analog dependence is imposed by the anisotropy field $H_\text{a}$, which monotonically decreases with an increasing temperature close to room temperature \cite{Kaack1999}. However, since the anisotropy fields are small, the temperature dependent contribution will be neglected in the following considerations.

Equation\,\ref{Eq1} also shows that a temperature induced change in the FMR frequency can be compensated experimentally by adjusting the external magnetic field. It is easy to see that an increase in temperature can be compensated by increasing $\mu_\text{0}H_{\text{ext}}$. The experimentally found temperature dependence of the compensation field is displayed in Fig.\,\ref{Fig2}(b). These data allow for a quantitative analysis of the function $M_\text{S}(T)$ using Eq.\,\ref{Eq1}. The change in $M_\text{S}$ as a function of temperature is assumed to be linear for the given temperature range and turns out to be $\Delta M_S/\Delta T = 0.38$\,kA/(m$\cdot$K) in agreement with previous reports\cite{Algra1982}.

\begin{figure}
	\includegraphics[viewport = 107 145 459 726, clip, scale=1.0, width=0.9\columnwidth]{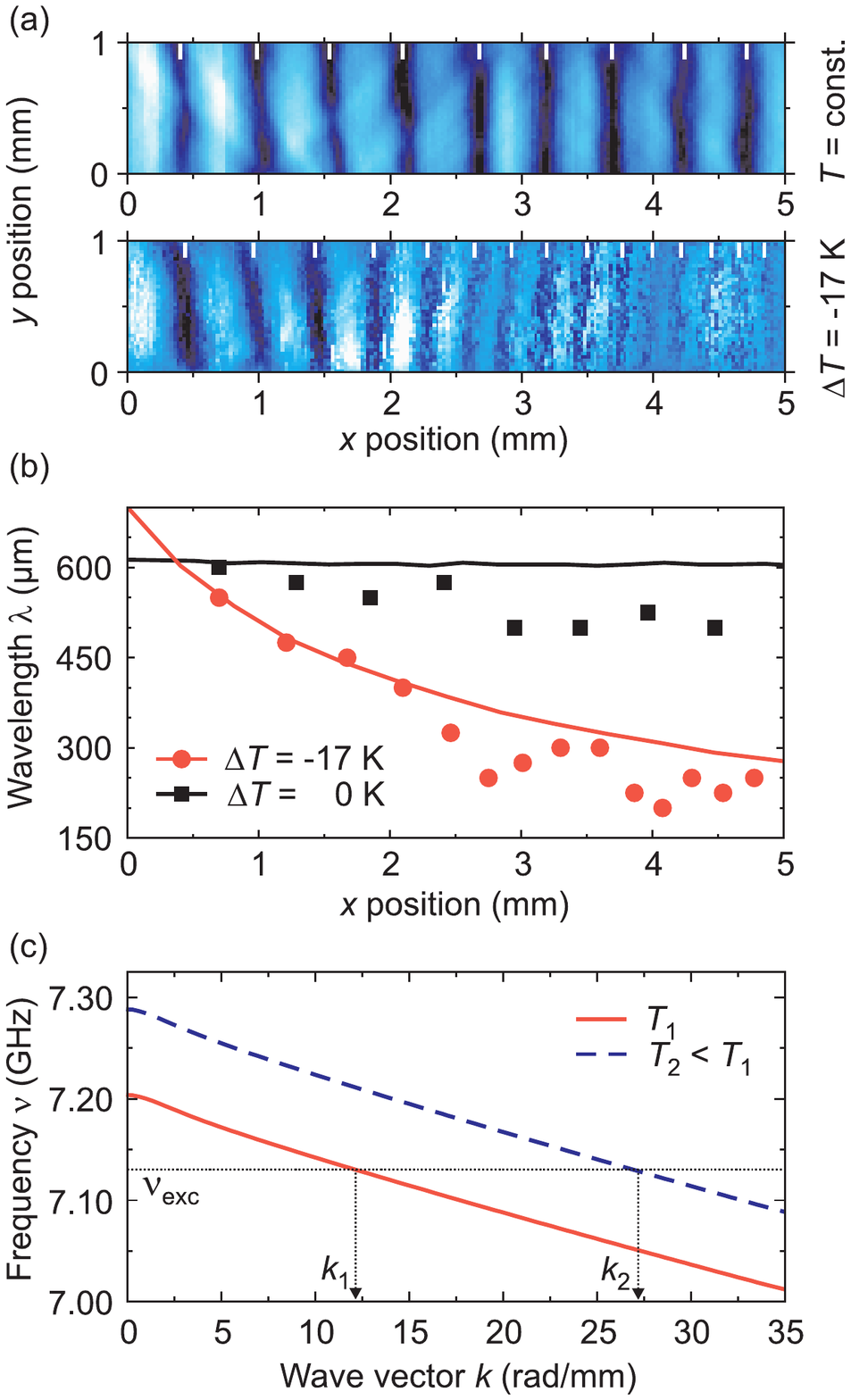}
	\caption{\label{Fig3} (Color online) (a) Phase-resolved Brillouin light scattering interference measurements revealing the wavelength reduction in a YIG waveguide with an applied thermal gradient (bottom) in comparison to uniform temperature (top). The interference between light that is scattered inelastically from spin waves and reference light with constant reference phase is measured as a function of the lateral position at the waveguide. A high (low) interference signal is indicated by white (black) color. Spin waves are excited at $x = 0$\,mm and propagate along the x direction. (b) Comparison between extracted values of the experimentally observed spin-wave wavelength (symbols) and calculated wavelength values based on the measured temperature along the stripe (lines) for uniform temperature (black) and an applied thermal gradient (pale red). (c) Calculated spin-wave dispersion relations for two different positions. The temperature induced frequency shift of the curves causes an increase in the spin-wave wave vector and hence a wavelength reduction.}
\end{figure}

In order to investigate the behavior of spin waves propagating in a region of inhomogeneous temperature, a thermal gradient has been applied to the waveguide so that spin waves are excited in a hot region near the antenna and propagate into colder regions of the YIG. This situation is shown in the thermal image of Fig.\,\ref{Fig1}(a). Using phase-resolved BLS spectroscopy\cite{Serga2006,Vogt2009} the intensity pattern of the interference between the spin-wave scattered light and reference light with a constant reference phase has been recorded, the latter being created by an electro-optic modulator driven by the common microwave generator. The interference minima and maxima correspond to the spin-wave scattered light being out of phase and in phase with the reference light, respectively. Mapping the spatial distribution of the interference signal allows for a visualization of the spin-wave wavelength, since the phase accumulation between two neighboring interference minima is $\Delta\Phi = 2\pi$. The resulting interference measurements for a waveguide with an applied thermal gradient and a reference measurement with a constant temperature are depicted in Fig.\,\ref{Fig3}(a). In both cases spin waves are excited at the antenna at $x = 0$\,mm. In the intensity graphs white (black) represents a high (low) interference signal. The interference minima are marked by white dashes. The upper panel of Fig.\,\ref{Fig3}(a) shows the results of spin waves propagating in a waveguide with homogeneous temperature, which has been measured for an external magnetic field of $\mu_0 H_\text{ext} = 180.6$\,mT. A constant spacing between the interference minima is observed indicating that the wavelength remains constant along the waveguide. In contrast, for spin waves propagating towards a colder region (lower panel of Fig.\,\ref{Fig3}(a)) a continuous reduction of the spacing with increasing distance from the antenna is detected. This has been achieved for a temperature difference of $\Delta T = -17\,$K along the waveguide section shown in Fig.~\,\ref{Fig3}(a). The magnetic field of $\mu_0 H_\text{ext} = 185.0$\,mT has been chosen to compensate the shift in the resonance spectrum at the position of the antenna and to keep the excitation conditions unchanged.

The data from Fig.\,\ref{Fig3}(a) indicate a wavelength reduction for spin waves propagating from a hot into a cold region. This can be understood by taking into account the change in saturation magnetization $M_\text{S}$ and hence the frequency shift of the dispersion relation. Figure\,\ref{Fig3}(c) shows the calculated dispersion relations for a position close to the excitation antenna (solid line) and at the end of the YIG waveguide\cite{comment} (dashed line) according to Ref.\,\onlinecite{Kalinikos1986}. Here, the dependence of $M_\text{S}(T)$ as determined above has been used. For a constant excitation frequency of $\nu_\text{exc} = 7.132$\,GHz a continuous reduction of the spin-wave wavelength with temperature is obtained. A comparison of the calculated wavelength values with the experimental data (Fig.\,\ref{Fig3}(b)) is in good agreement. 
For increasing distance from the antenna the experimental results exhibit an increasing deviation to smaller wavelength values even for the case of a constant sample temperature. This deviation has a large extension through the sample, and thus cannot be caused by shape-induced variations of the demagnetizing field, which are most pronounced near the sample ends. Since this wavelength reduction appears in both measurements, it might be caused by an intrinsic effect of the spin waves, i.e. the increase in $M_\text{S}$ due to the decrease of the spin-wave intensity in the course of the wave propagation away from the antenna. A similar effect has been reported for phase measurements of intensive BVMSW packets\cite{Schneider2007}. Theoretical approaches to describe this intrinsic effect by accurately considering the long-range interactions of the dynamic magnetization on the internal field distribution exist in literature already\cite{Puszkarski2007}, however, a quantitative analysis is beyond the scope of this Letter.

\begin{figure}
	\includegraphics[viewport = 119 79 477 543, clip, scale=1.0, width=0.9\columnwidth]{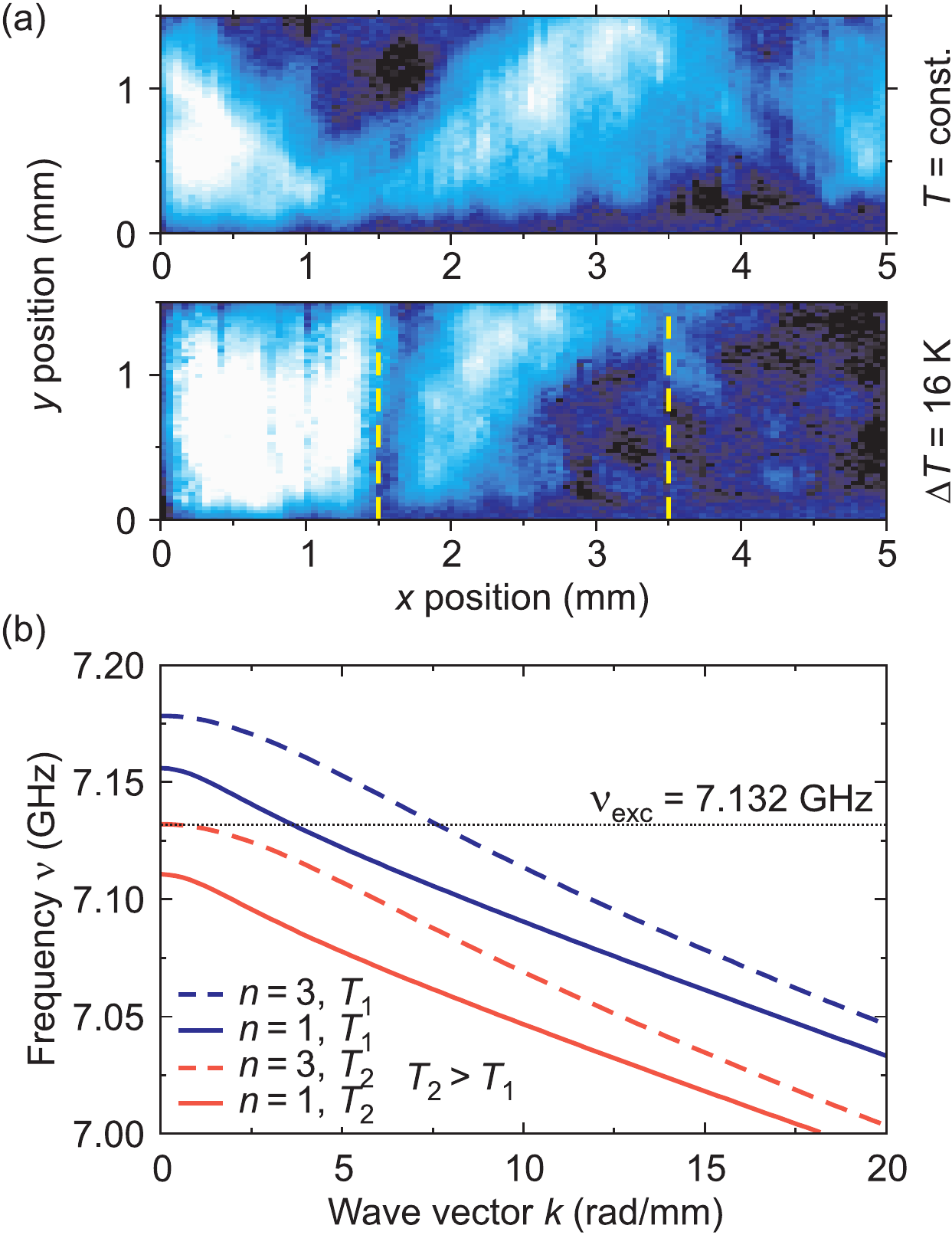}
	\caption{\label{Fig4} (Color online) (a) Brillouin light scattering measurements of the spin-wave reflection in a thermal gradient. The spatial spin-wave intensity distribution in a YIG waveguide is shown for uniform temperature (top) and increasing temperature (bottom) along the spin-wave propagation direction. White (black) color indicates high (low) spin-wave intensity. Spin waves are excited at $x = 0$\,mm and propagate along the x direction. (b) Calculated spin-wave dispersion relations for a position near the antenna (dark blue) and at a distance of 3.5\,mm to the antenna (pale red) illustrating the underlying mechanism of the spin-wave reflection.}
\end{figure}

A further experiment of spin waves traveling towards higher temperatures has been realized in the configuration shown by the thermal image in Fig.\,\ref{Fig1}(b). Figure~\ref{Fig4}(a) shows BLS measurements of the spin-wave intensity distribution in the YIG waveguide with a magnetic field of $\mu_0 H_\text{ext} = 178.6$\,mT. By selecting these settings it can be ensured that the microwave excitation frequency of $\nu_\text{exc} = 7.132$\,GHz is close to the FMR frequency of the waveguide and that the applied temperature gradient in the waveguide is high enough to compensate the intrinsic wavelength transformation, which now acts opposite to the externally imposed wave vector modification. In a reference measurement with uniform waveguide temperature (upper panel of Fig.\,\ref{Fig4}(a)) the spin-wave intensity distribution resembles a snake-like pattern which is understood to be the superposition of several transversal waveguide modes\cite{Bauer1998}. The present case can be attributed to an interference of the two modes $n = 1$ and $n = 3$, where $n$ denotes the number of antinodes along the short axis of the waveguide. In addition, the propagation is influenced by a caustic behavior to a small extent\cite{Schneider2010}. Measurements of the spin-wave intensity distribution in a waveguide with a temperature difference of $\Delta T = 16$\,K (lower panel of Fig.\,\ref{Fig4}(a)) show a drastically reduced intensity in the regions far from the antenna (beyond dashed line at $x = 3.5$\,mm) while there is an increase in the detected signal close to the excitation region (up to dashed line at $x = 1.5$\,mm), both indicating a reflection of the spin waves in the thermal gradient.

The reflection can be understood in analogy to the considerations above. Increasing temperature causes a downward shift of the dispersion to lower frequencies. The effect on the spin-wave propagation can be seen from the calculations of the dispersion relation (Fig.\,\ref{Fig4}(b)) for a position near the antenna (dark blue lines) and for a position at $x = 3.5$\,mm with a higher temperature (pale red lines). Upon propagation to larger distances from the antenna the temperature difference will induce a dispersion shift below the excitation line ($\nu_\text{exc} = 7.132$\,GHz). Thus, spin waves with the given frequency can no longer exist and have to be reflected. The temperature gradient causes the formation of a `forbidden' region for these spin waves. Since the dispersion relations of the two waveguide modes $n = 1$ (solid lines) and $n = 3$ (dashed lines) are displaced in frequency, their critical points where reflection occurs do not coincide. Using the temperature information from Fig.\,\ref{Fig1}(b) to calculate the local values of $M_\text{S}$ allows for a determination of the theoretically expected reflection points of $x = 1.3$\,mm for the transversal waveguide mode $n = 1$ as well as $x = 3.1$\,mm for $n = 3$. These values match the two experimentally determined positions indicated by the dashed lines in Fig.\,\ref{Fig4}(a), where the intensity pattern changes.

In conclusion, the results presented demonstrate that it is possible to effectively manipulate spin waves by the application of a thermal gradient. With their propagation into regions of lower temperature a reduction of the spin-wave wavelength to half of its original value is observed. Moreover, spin waves can experience a reflection upon traveling into regions of higher temperatures. Thus, thermal gradients are a valuable tool for influencing the propagation properties of spin waves. There is much potential in temperature effects, which needs to be explored and exploited for application in spintronic or magnon logic devices.

%\section{}
%\label{}
%\subsection{}
%\subsubsection{}

% If in two-column mode, this environment will change to single-column format so that long equations can be displayed. 
%\begin{widetext}
%$$\mbox{put long equation here}$$
%\end{widetext}

% Figures should be put into the text as floats. 
% Use the graphics or graphicx packages (distributed with LaTeX2e).
% Put the label that you will use with \ref{} command in the braces of the \label{} command.
%
% \begin{figure}
% \end{figure}

% \begin{table}
% \end{tabular}
% \end{table}

The authors would like to thank E. Saitoh for fruitful discussions. Financial support by the Deutsche Forschungsgemeinschaft (DFG, VA 735/1-1) within Priority Program 1538 ``Spin Caloric Transport'' is gratefully acknowledged.
% Put your acknowledgments here.



\begin{thebibliography}{1}

\bibitem{Bauer2012}G.\,E.\,W.~Bauer, E.~Saitoh, and B.\,J.~van~Wees, Nat. Mat. \textbf{11}, 391 (2012).

\bibitem{Uchida2010}K.~Uchida, J.~Xiao, H.~Adachi, J.~Ohe, S.~Takahashi, J.~Ieda, T.~Ota, Y.~Kajiwara, H.~Umezawa, H.~Kawai, G.\,E.\,W.~Bauer, S.~Maekawa, and E.~Saitoh, Nat. Mat. \textbf{9}, 894 (2010).

\bibitem{Hirsch1999}J.\,E.~Hirsch, Phys. Rev. Lett. \textbf{83}, 1834 (1999).

\bibitem{Uchida2010a}K.~Uchida, H.~Adachi, T.~Ota, H.~Nakayama, S.~Maekawa, and E.~Saitoh, Appl. Phys. Lett. \textbf{97}, 172505 (2010).

\bibitem{Xiao2010}J.~Xiao, G.\,E.\,W.~Bauer, K.~Uchida, E.~Saitoh, and S.~Maekawa, Phys. Rev. B \textbf{81}, 214418 (2010).

\bibitem{Damon1961}R.\,W.~Damon and J.\,R.~Eshbach, J. Phys. Chem. Solids \textbf{19}, 308 (1961).

\bibitem{Hillebrands2000}B.~Hillebrands, in {\it Light Scattering in Solids VII}, edited by M.~Cardona and G.~G\"{u}ntherodt (Springer Berlin/Heidelberg, 2000) pp.174-289.

\bibitem{Kittel1948}C.~Kittel, Phys. Rev. \textbf{73}, 155 (1948).

\bibitem{Gieniusz1987}R.~Gieniusz and L.~Smoczy\'{n}ski, J. Magn. Magn. Mater. \textbf{66}, 366 (1987).

\bibitem{Kaack1999}M.~Kaack, S.~Jun, S.\,A.~Nikitov, and J.~Pelzl, J. Magn. Magn. Mater. \textbf{204}, 90 (1999).

\bibitem{Hansen1974}P.~Hansen, P.~R\"{o}schmann, and W.~Tolksdorf, J. Appl. Phys. \textbf{45}, 2728 (1974).

\bibitem{Roeschmann1981}P.~R\"{o}schmann and P.~Hansen, J. Appl. Phys. \textbf{52}, 6257 (1981).

\bibitem{Gurevich1996}A.\,G.~Gurevich and G.\,A.~Melkov, {\it Magnetization Oscillations and Waves} (CRC Press, Boca Raton, 1996) p. 202 ff.

\bibitem{Lvov1993}V.~Cherepanov, I.~Kolokolov, and V.~L'vov, Phys. Rep. \textbf{229}, 81 (1993).

\bibitem{Algra1982}H.\,A.~Algra and P.~Hansen, Appl. Phys. A \textbf{29}, 83 (1982).

\bibitem{Serga2006}A.\,A.~Serga, T.~Schneider, B.~Hillebrands, S.\,O.~Demokritov, and M.\,P.~Kostylev, Appl. Phys. Lett. \textbf{89}, 063506 (2006).

\bibitem{Vogt2009}K.~Vogt, H.~Schultheiss, S.\,J.~Hermsdoerfer, P.~Pirro, A.\,A.~Serga, and B.~Hillebrands, Appl. Phys. Lett. \textbf{95}, 182508 (2009).

\bibitem{comment}Experimental parameters: YIG stripe, width $w = 1.5$\,mm, thickness $t = 6.7\,\mu$m, $M_\text{S} = 139.3$\,kA/m (at RT); gyromagnetic ratio $\gamma = 28$\,GHz/T; magnetic field $\mu_0 H_\text{ext} = 182.3$\,mT.

\bibitem{Kalinikos1986}B.\,A.~Kalinikos and A.\,N.~Slavin, J. Phys. C: Solid State Phys. \textbf{19}, 7013 (1986).

\bibitem{Schneider2007}T.~Schneider, A.\,A.~Serga, B.~Hillebrands, and M.\,P.~Kostylev, Europhys. Lett. \textbf{77}, 57002 (2007).

\bibitem{Puszkarski2007}H.~Puszkarski, M.~Krawczyk, and J.-C.\,S.~L\'{e}vy, J. Appl. Phys. \textbf{101}, 024326 (2007).

\bibitem{Bauer1998}O.~B\"{u}ttner, M.~Bauer, C.~Mathieu, S.\,O.~Demokritov, B.~Hillebrands, P.\,A.~Kolodin, M.\,P.~Kostylev, S.~Sure, H.~D\"{o}tsch, V.~Grimalsky, Yu.~Rapoport, and A.\,N.~Slavin, IEEE Trans. Magn. \textbf{34}, 1381 (1998).

\bibitem{Schneider2010}T.~Schneider, A.\,A.~Serga, A.\,V.~Chumak, C.\,W.~Sandweg, S.~Trudel, S.~Wolff, M.\,P.~Kostylev, V.\,S.~Tiberkevich, A.\,N.~Slavin, and B.~Hillebrands, Phys. Rev. Lett. \textbf{104}, 197203 (2010).

\end{thebibliography}
\end{document}